\documentclass{jetpl}
\usepackage{graphicx}
\twocolumn

\lat
\def\e{{\rm e}}

\newcommand{\be}{\begin{equation}}
\newcommand{\ee}{\end{equation}}
\newcommand{\bea}{\begin{eqnarray}}
\newcommand{\eea}{\end{eqnarray}}
\newcommand{\bg}{\begin{gather}}
\newcommand{\eg}{\end{gather}}
\newcommand{\bseq}{\begin{subequations}}
\newcommand{\eseq}{\end{subequations}}

\renewcommand{\ch}{\mathop{\rm ch}\nolimits}
\newcommand{\sh}{\mathop{\rm sh}\nolimits}
\renewcommand{\ln}{\mathop{\rm ln}\nolimits}

\renewcommand{\Im}{\mathop{\rm Im}\nolimits}

\newcommand{\bra}[1]{\langle #1 |}
\newcommand{\ket}[1]{| #1 \rangle}

\title{Real--Time Instantons and Suppression of Collision--Induced Tunneling}

\rtitle{Real--Time Instantons and Suppression of Collision--Induced Tunneling}

\sodtitle{
Real--Time Instantons and Suppression of Collision--Induced Tunneling
}

\author{D.~Levkov$^{\;a,b}$\/\thanks{e-mail: levkov@ms2.inr.ac.ru},
S.~Sibiryakov$^{\; a}$}

\rauthor{D.~Levkov, S.~Sibiryakov}

\sodauthor{D.~Levkov, S.~Sibiryakov}

\address{
$^a$ Institute for Nuclear Research of the Russian Academy of
Sciences,\\ 60th October Anniversary prospect 7a, Moscow 117312, Russia\\
$^b$ Moscow State University, Department of Physics,
    Vorobjevy Gory, Moscow, 119899, Russia
}

\abstract{
We consider tunneling processes in QFT induced by collisions of
elementary particles. We propose a semiclassical method for estimating
the probability of these processes in the limit of very high
collision energy. As an illustration, we evaluate the maximum
probability of induced tunneling between
different vacua in a $(1+1)$--dimensional scalar model with boundary
interaction.
}


\begin{document}
\maketitle

In many models of field theory one encounters tunneling transitions
between states separated by an energy barrier of  finite height
$E_S$, the famous examples are false vacuum decay in scalar
theories~\cite{Coleman:1977py} and topology--changing transitions in
gauge--Higgs theory~\cite{Belavin:1975fg,Klinkhamer:1984di}. In the
weak coupling regime, the rate of tunneling at zero energy is exponentially 
small~\cite{Coleman:1977py,Belavin:1975fg}, but one suspects 
the suppression to vanish once energy exceeding the
height of the barrier is injected into the system. A particular
way of inducing the tunneling process is a  collision of two
highly energetic particles. It was conjectured
some time ago~\cite{Ringwald:1990ee}
that the collision--induced tunneling processes may become 
unsuppressed at high collision energies.
Semiclassical study of scalar and gauge--Higgs 
theories~\cite{Kuznetsov:1997az,Bezrukov:2003er} showed, however, that this is 
not the case: the tunneling probability remains exponentially 
small even if the collision energy $E$ exceeds considerably the
barrier height $E_S$. 
Furthermore, analyses of toy
models~\cite{Voloshin:1993dk,Rubakov:1994hz} and unitarity 
arguments~\cite{Zakharov:1990xt} suggest that the collision--induced 
transitions should remain exponentially suppressed even if $E$ tends
to infinity. In other words, it was proposed that, contrary to the
initial conjecture, the probability of the process has the form,
\begin{equation}
\label{PE}
{\cal P}(E) \propto \mathrm{e}^{-F(E)/g^2}\;,
\end{equation}
where $g^2$ is a small coupling constant, and the suppression exponent
$F(E)$ remains positive at all energies.
This has been confirmed recently by direct 
calculation~\cite{Levkov:2004tf} of the suppression exponent
in the whole range of energies in a toy
two--dimensional model. In addition, it was found
in~\cite{Levkov:2004tf}  that
the suppression exponent reaches its minimum $F_{\rm m}$ 
at certain optimal energy
$E_{\rm o}$ and remains constant above this energy,
\begin{equation}
\label{limiting}
F(E)=F_{\rm m}~,~~~E>E_{\rm o}\;.
\end{equation}
One may ask whether or not the
formulas~\eqref{PE},~\eqref{limiting} are valid for other models,
with model--dependent values of $F_\mathrm{m}$, $E_\mathrm{o}$.

In this letter we give a general semiclassical method for
evaluating the minimum value $F_{\rm m}$ of the suppression exponent 
and
the energy $E_{\rm o}$  at which this minimum is achieved. 
Our procedure is
essentially an adaptation of the method of 
Ref.~\cite{Rubakov:1992ec}, but it is more straightforward
and technically simpler.
We consider the case in which, after  
the appropriate rescaling of the fields, the action $S$ takes
the form $S= \tilde{S}/g^2$, where $\tilde{S}$ does not explicitly
depend on the small coupling constant $g$. One observes that 
$g^2$ plays effectively the role of the Planck 
constant, and the limit $g\to0$ which we consider below, corresponds 
to a semiclassical situation. 

Our starting point is 
the inclusive probability of tunneling
from states with a given number of incoming
particles and {\em any} energy,
\begin{equation}
\label{PN}
{\cal P}_\mathrm{m} (N) = \sum\limits_{i,f} |\langle f| 
\hat{\cal  U}(T_f,T_i) \hat{P}_N 
|i\rangle|^2\;.
\end{equation}
Here $\hat{\cal U}(T_f,T_i)$ is the evolution operator, 
and $\hat{P}_N$ denotes the projector onto
states with $N$ particles. The initial and final states,
$|i\rangle$ and $|f\rangle$, respectively, are at the  
different sides of the potential barrier. Here and below the
limit $T_f\to +\infty$, $T_i\to -\infty$ is assumed. 
As we will shortly see, 
the quantity (\ref{PN}) can be evaluated 
semiclassically provided  the initial 
number of particles is parametrically large, 
$N=\tilde{N}/g^2$. We will 
see that the result
has a typical exponential form,
\begin{equation}
\label{PNresult}
{\cal P}_\mathrm{m}(N) \propto \mathrm{e}^{-F_\mathrm{m}(\tilde{N})/g^2}\;.
\end{equation}
We use the result~\eqref{PNresult} as a source of information
on the probability of collision--induced tunneling. 
It is clear that the inclusive multiparticle probability 
${\cal P}_\mathrm{m}(N)$ sets an upper bound on 
the two-particle probability of interest,
${\cal P}(E)$, at arbitrarily high  energies $E$. Indeed, 
the energy of the initial state in (\ref{PN}) can be arbitrarily high,
while any 
initial two--particle state can be promoted to the multiparticle one by adding 
a number of ``spectator'' particles which do not interfere with the 
tunneling process.  Hence, the exponential suppression of 
${\cal P}_\mathrm{m}(N)$ entails the
exponential suppression of ${\cal P}(E)$,
and the inequality 
\begin{equation}
F_{\rm m}(\tilde{N})\leq F_{\rm m}
\label{FleqF}
\end{equation}
holds.
Following Ref.~\cite{Rubakov:1992ec}, we conjecture 
that 
\begin{equation}
\label{lim}
\lim\limits_{\tilde{N}\to0} F_\mathrm{m}(\tilde{N})=F_{\rm m}\;.
\end{equation}
The conjecture~\eqref{lim} is based on the observation that in the leading
semiclassical approximation the probability of tunneling does not
depend on the details of the initial state provided the initial number of
particles is much smaller than $1/g^2$.  The 
analogs of the formula~\eqref{lim} have been checked in
various situations~\cite{Tinyakov:1991fn,Bonini:1999kj,Bezrukov:2003tg}.

Now, we proceed
to the semiclassical evaluation of the multiparticle exponent 
$F_\mathrm{m}(\tilde{N})$. Our strategy is to represent the 
inclusive probability ${\cal P}_\mathrm{m}(N)$ 
in the form of  path integral and evaluate the latter
making use of the saddle--point technique.  
Taking into account that 
$\hat{P}_N\cdot\hat{P}_N = \hat{P}_N$, one recasts Eq.~\eqref{PN} in the 
form,
\begin{equation}
\label{P1}
\begin{split}
{\cal P}_\mathrm{m}(N) =& 
\sum\limits_{i',i'',f}\langle f|\hat{\cal U}|i'\rangle
\langle i'|\hat{P}_N|i''\rangle\langle i''|\hat{\cal U}|f\rangle\\
=&\int{\cal D}[\phi_f,a,a^*,b,b^*] \e^{-\frac{1}{g^2}
\int d {\bf k}[a_{\bf k} a_{\bf k}^* + b_{\bf k} b_{\bf k}^*]}\\
&~~~~~\times 
\bra{\phi_f}\hat{\cal U}\ket{a}\; \bra{a}\hat{P}_N\ket{b}\;
[\bra{\phi_f}\hat{\cal U}\ket{b}]^*\;,
\end{split}
\end{equation}
where 
$\ket{\phi_f}$ are eigenstates of the field operators which we
denote collectively by $\hat{\phi}$,
while $\ket{a}$, $\ket{b}$ are the coherent states,
\[
\hat{\phi}({\bf x})\ket{\phi}=\frac{\phi({\bf x})}{g}\ket{\phi}~,~~~
\hat{a}_{\bf k}\ket{a}=\frac{a_{\bf k}}{g}\ket{a}\;.
\]
Hereafter we use the shorthand 
notation $\hat{\cal U} \equiv \hat{\cal U}(T_f,T_i)$.
Making use of the standard path integral for the transition
amplitude in $\phi$-representation, one writes
\begin{equation}
\label{Smatr}
\langle \phi_f|\hat{\cal U}|a\rangle = 
\!\!\!\int\limits_{\phi(T_f) = \phi_f}\!\!\!{\cal D} \phi
\exp\left\{\frac{1}{g^2}\big(i\tilde{S}[\phi] + 
B_i(\phi_i,a)\big)\right\},
\end{equation}
where the boundary term $B_i$ comes from 
the initial matrix element $\langle\phi\ket{a}$,
\begin{multline}
\nonumber
B_i(\phi_{i},a) = \int d{\bf k} 
\left\{-\frac{1}{2} a_{\bf k} a_{-{\bf k}}
   \right.\\\left. -
\frac{\omega_{\bf k}}{2} \phi_{i}({\bf k}) \phi_{i}(-{\bf k}) + 
\sqrt{2\omega_{\bf k}} a_{\bf k} 
  \phi_{i}({\bf k})\right\}\;.
\end{multline}
In this expression $\phi_{i}({\bf k})$ stands for the 
spatial Fourier transform of the fields at $t=T_{i}$.
For the matrix element of the projector 
$\hat{P}_N$ one obtains (see e.g.~\cite{Tinyakov:1992dr,Rebbi:1996qt})
\begin{equation}
\label{proj}
\langle a|\hat{P}_N | b\rangle = \!\!\int\limits_{-i\infty}^{i\infty} 
\!\!\! d\theta
\exp\left\{\frac{1}{g^2}\left(\tilde{N}\theta + 
\int d{\bf k} \, a_{\bf k}^* b_{\bf k} \mathrm{e}^{-\theta}\right)\right\}\;.
\end{equation}
Substituting Eqs.~\eqref{Smatr},~\eqref{proj} into the expression~\eqref{P1}
and performing integration over the variables $b$, $b^*$, one obtains 
the desired path integral representation,
\begin{equation}
\label{P3}
{\cal P}_\mathrm{m}(N) = \int\limits_{\phi(T_f)=\phi'(T_f)}
{\cal D}[\phi,\phi',a,a^*] d\theta \; 
\mathrm{e}^{-F/g^2}\;,
\end{equation}
where
\begin{multline}
\label{Fm}
F= -\tilde{N}\theta - i\tilde{S}[\phi] + i\tilde{S}[\phi'] \\- B_i(\phi_i,a) 
- B_i^*(\phi_i',a)
+ \int d{\bf k} \; a_{\bf k}^* a_{\bf k} \mathrm{e}^{\theta} \;.
\end{multline}
Note that the integration over $\phi'$ in \eqref{P3} comes from 
the path integral representation for the complex conjugate amplitude
$[\langle\phi_f|\hat{\cal U}|b\rangle]^*$.

The functional $F$ defined in Eq.~(\ref{Fm}) is independent of the
coupling constant $g$. Hence, at weak coupling the integral
(\ref{P3}) is saturated by its saddle point.  The
saddle--point equations are as follows.  
Extremization with respect to $\phi$ and $\phi'$ gives the
classical field equations
\begin{subequations}
\label{eq}
\begin{equation}
\label{eqt}
\frac{\delta \tilde{S}}{\delta \phi} = 
\frac{\delta \tilde{S}}{\delta \phi'} = 0\;.
\end{equation}
Boundary conditions for these equations are obtained by varying  the 
expression~\eqref{Fm} with 
respect to the initial and final values of the fields. 
Using the relation $\delta S/\delta \phi(T_f,{\bf x}) =
\dot{\phi}(T_f, {\bf x})$ and taking into account the constraint
(see. Eq.(\ref{P3}))
\be
\label{eqf1}
\phi(T_f, {\bf x})=\phi'(T_f,{\bf x})\;,
\ee
one obtains,
\begin{equation}
\label{eqf}
\dot\phi(T_f,{\bf x})=\dot\phi'(T_f,{\bf x})\;.
\end{equation}
In the initial 
asymptotic region $T_i\to -\infty$ the evolution of the fields 
$\phi$, $\phi'$ is
linear, and one writes\footnote{For concreteness we assume that the
initial state is an excitation above the vacuum $\phi=0$.}
\begin{align}
\nonumber
&\phi_i = \int \frac{d {\bf k}}{(2\pi)^{1/2}\sqrt{2\omega_{\bf k}}} 
  \left(
    f_{\bf k}\mathrm{e}^{-i\omega_{\bf k} T_i} + 
    g_{-{\bf k}}^*\mathrm{e}^{i\omega_{\bf k} T_i}\right)
    \e^{i{\bf k}{\bf x}}\;,
\end{align}

\noindent
\begin{align}
\nonumber
&\phi_i' =\! \int \frac{d {\bf k}}{(2\pi)^{1/2}\sqrt{2\omega_{\bf k}}} 
  \left(
    {f'_{\bf k}}\mathrm{e}^{-i\omega_{\bf k} T_i} + 
    {g'^*_{-{\bf k}}}\mathrm{e}^{i\omega_{\bf k} T_i}
\right)\e^{i{\bf k}{\bf x}}\;.
\end{align}
The variation of the functional $F$ with respect to $\phi_i$, $\phi_i'$, 
$a$, $a^*$ yields the following relations between the frequency
components,
\begin{equation}
\label{eqi}
f'_{\bf k} = f_{\bf k} \mathrm{e}^{\theta}\;,\;\; 
{g'^*_{\bf k}} = g_{\bf k}^* \mathrm{e}^{-\theta}\;.
\end{equation}
\end{subequations}
The set of saddle--point equations~\eqref{eq} can
be simplified if we recall that the configurations $\phi$ and $\phi'$
saturate the amplitude and its complex conjugate, respectively. 
This suggests the Ansatz
$\phi'(t,{\bf x}) = [\phi(t,{\bf x})]^*, $
which is compatible with the boundary value problem~\eqref{eq}
provided  the saddle--point value of $\theta$ is real. 
Then, the boundary value problem is 
formulated  in terms of a single set of fields $\phi(t,{\bf x})$. 
The conditions (\ref{eqf1}), \eqref{eqf} 
imply the  reality of the fields in the asymptotic
future,
\begin{subequations}
\label{BC}
\begin{equation}
\label{BCF}
\Im\phi(t,{\bf x}) \to 0\,, \;\;
\Im\dot\phi(t,{\bf x}) \to 0 \;\; \mathrm{as} \;\; t\to+\infty\;,
\end{equation}
while Eqs.~\eqref{eqi} read
\begin{equation}
\label{BCI}
f_{\bf k} = \mathrm{e}^{-\theta} g_{\bf k}\;.
\end{equation}
\end{subequations}
The boundary condition in the asymptotic past, Eq.~\eqref{BCI},
can be understood as follows. In the limit $\theta\to +\infty$ it coincides 
with the Feynman boundary condition and thus corresponds to the initial state 
with semiclassically small number of particles, $\tilde{N}\to 0$;
finite $\theta$ picks up the most favourable state with non--zero 
$\tilde{N}$. 

The number of equations in the boundary value problem (\ref{eqt}), (\ref{BC})
is equal to the
number of unknowns.  
Generically, for a given value of $\theta$ this problem has a unique 
solution $\phi_{rt}(t,{\bf x})$. We call this solution  
``real--time instanton'', as it lives on the real time axis, 
in contrast to the ordinary 
instanton which is defined in Euclidean time. Note that the boundary
condition (\ref{BCI}) implies that the real--time instanton is
complex--valued. On the other hand, its imaginary part should vanish in the
asymptotic future due to the condition (\ref{BCF}). 
Let us discuss the consequences of this property. 
Assume that at large (but finite) times the solution gets linearized 
about some real static configuration, ${\phi_{rt}(t,{\bf x}) = \phi_s({\bf x}) 
+ \delta\phi(t,{\bf x})}$.
Then Eq.(12a) implies that $\Im \delta\phi \to 0$ as
$t\to +\infty$. 
This amounts to requiring that the configuration $\phi_s$ is unstable,
so that $\Im \delta\phi$
evolves along its negative mode,
$\Im\delta\phi \propto \e^{-{\rm const}\cdot t}$. 
The natural
candidate for $\phi_s({\bf x})$ is 
the static solution ``sitting'' on top of the potential 
barrier separating the sectors of
initial and final states. 
%
Accepting the terminology of gauge-Higgs 
theories 
\cite{Klinkhamer:1984di} we call this
solution ``sphaleron''\footnote{In case of scalar theories such
  solution
 is known as ``critical bubble''~\cite{Coleman:1977py}.}.
We arrive at the conclusion that the real--time instantons describe
formation ~of~ the ~sphaleron as ~~${t\to +\infty}$.

\noindent
This is in fact 
a common property of solutions relevant for collision--induced
tunneling at energies higher than $E_S$ 
\cite{Bezrukov:2003tg,Bezrukov:2003er,Levkov:2004tf}.
The transition is completed by the decay of the sphaleron into
the states of interest, which
proceeds
with probability of order one. 

The last saddle--point equation obtained by varying  the
functional (\ref{Fm}) 
with respect 
to the parameter $\theta$, 
relates the value of $\theta$ to the initial number
of particles,
\begin{equation}
\label{N}
\tilde{N} = \int d{\bf k} f_{\bf k} g_{\bf k}^*\;.
\end{equation}
Substituting 
the solution into Eq.~\eqref{Fm}, one obtains the 
formula (\ref{PNresult}), with the suppression exponent 
\begin{equation}
\label{Fmresult}
F_\mathrm{m}(\tilde{N}) = -\tilde{N}\theta + 
2\Im\left\{\!\tilde{S}[\phi_{rt}] 
+\frac{1}{2}\! \int \!\!d{\bf x} \phi_{rt}\dot\phi_{rt} 
\bigg|_{t=T_i}\right\}.
\end{equation}
Note that the term in braces is the action of the real--time
instanton integrated by parts with respect to time. 

At finite value of the parameter $\theta$ (non-zero value of
$\tilde{N}$) the real--time instanton is, by construction, a smooth
solution to the equations of motion and has a well--defined
classical energy $E_{\rm o}(\tilde{N})$. At $t\to +\infty$ the real--time
instanton describes semiclassical evolution in the final state.
Thus, $E_{\rm o}(\tilde{N})$ coincides with 
the energy of the final state saturating the 
probability~(\ref{PN}).
This means that if we
restrict the  sum in~\eqref{PN} to the states with the {\em
  fixed} energy $E = E_{\rm o}(\tilde{N})$ (cf. Ref.~\cite{Rubakov:1992ec})
we obtain the same result~(\ref{PNresult}). One concludes that
$E_{\rm o}(\tilde{N})$ is the optimal energy for tunneling from the states
with given number of particles $N = \tilde{N}/g^2$.
The limit 
\begin{equation}
E_{\rm o}=\lim\limits_{\tilde{N}\to 0}E_{\rm o}(\tilde{N})
\label{EleqElN}
\end{equation}
determines the optimal
energy for collision--induced tunneling\footnote{In general, we cannot
exclude the situation when $E_{\rm o}(\tilde{N})\to \infty$ as 
$\tilde{N}\to0$. In the example
below, $E_{\rm o}$ is finite.}. It is easy to
understand the most favourable transition  
at higher energies. The system releases the
energy excess $(E-E_{\rm o})$ by a perturbative emission of a few particles
(which costs only a power suppression in $g^2$) and tunnels at
the optimal energy $E_{\rm o}$ \cite{Voloshin:1993dk,Levkov:2004tf}. 
Due to this process $F(E)$ stays constant 
at $E > E_{\rm o}$, i.e. the formula~\eqref{limiting} holds.

Let us illustrate our 
method by considering a simple example. We consider 
free massless scalar field $\phi(t,x)$ living in
(1+1) dimensions on a half--line $x>0$, with self--interaction 
localized at the 
boundary point $x=0$. The action of the model is\footnote{
Normally, one should use some infrared regularization for the massless scalar
model in (1+1) dimensions.
However, the specifics of regularization turn out to be irrellevant for
our purposes. }
\begin{equation}
\label{L}
S = \frac{1}{g^2}\!\int\! dt
\left\{\frac{1}{2}\int\limits_0^\infty\! dx (\partial_\mu \phi)^2 
- \mu[1-\cos(\phi(t,0))]\right\}\;,
\end{equation}
where the parameter $\mu$ sets the 
characteristic
energy scale of the boundary interaction. The 
model~\eqref{L} is used in solid state physics to
describe the transport in quantum wires~\cite{Kane} and Josephson
chains with defects~\cite{Fazio}. A detailed semiclassical
treatment of the model (\ref{L}) is given in \cite{Levkov:2004tf}. 

The model (\ref{L}) has
a number of vacua $\phi_n = 2\pi n$, $n=0,\pm 1, \dots$, which are 
separated by the potential barriers of the height $E_S = 2\mu/g^2$ 
determined by 
the maximum of the boundary potential. One also finds an unstable static 
solution $\phi_s = \pi$, sphaleron, 
which ``sits'' on the top of the ``first'' 
potential barrier. 
The process we are interested in is tunneling 
between the vacua $\phi = 0$ and $\phi = 2\pi$ induced by a highly energetic
particle scattering off the boundary. To calculate the minimum suppression
exponent and the optimal energy of 
this collision--induced tunneling process, one finds the 
family of real--time instantons by
solving the boundary value problem~\eqref{eqt},~\eqref{BC}.

Since the bulk evolution of $\phi(t,x)$ is that of free massless 
scalar field, we represent the general solution in the form
\begin{equation}
\label{decomp}
\phi(t,x) = \varphi_{\mathrm{in}}(t+x) + \varphi_{\mathrm{out}}(t-x)\;,
\end{equation}
where $\varphi_{\mathrm{in}}$ and $\varphi_{\mathrm{out}}$ 
are the incoming and outgoing waves. 
The boundary interaction leads to non--linear equation at $x=0$,
\begin{equation}
\label{BCX1}
\partial_x \phi = \mu \sin{\phi}, \;\; x=0\;.
\end{equation}
Due to the 
condition~\eqref{BCF}, the outgoing wave $\varphi_{\mathrm{out}}$ is real. 
Introducing the real and imaginary parts of the incoming wave, 
$\varphi_{\mathrm{in}}(\xi) = a(\xi)+ib(\xi)$, 
one rewrites Eq.~\eqref{BCX1} as a set of two real
equations,
\begin{align}
\label{b}
&b' = \mu\sh{b}\cos{u}\;,\\
\label{c}
&u' = 2a' - \mu\ch{b}\sin{u}\;,
\end{align}
where $u(\xi) = a(\xi) + \varphi_{\mathrm{out}}(\xi)$. 
The remaining conditions in the asymptotic 
past, Eq.~\eqref{BCI}, should be imposed on the frequency components of the 
incoming wave $\varphi_{\mathrm{in}}$. 
To this end, one performs Fourier expansion of 
$\varphi_{\mathrm{in}}$,
\begin{equation}
\label{Four}
\varphi_{\mathrm{in}}(t+x) = \int dk\; \varphi_{\mathrm{in}}(k) 
\mathrm{e}^{ik(t+x)}\;,
\end{equation}
and finds that the positive and negative frequency components 
of the solution, $f_{-k}$ and
$g_{-k}^*$, are proportional to $\varphi_{\mathrm{in}}(-k)$ and 
$\varphi_{\mathrm{in}}(k)$, $k>0$,
respectively. Thus, Eq.~\eqref{BCI} takes the form 
$\varphi_{\mathrm{in}}(-k) = \mathrm{e}^{-\theta} 
[\varphi_{\mathrm{in}}(k)]^*$, where $k>0$.
It is straigforward to check using the Cauchy formula that the latter
condition is equivalent in its turn to the following relation between the real
and imaginary parts of the initial wave (see Ref.~\cite{Levkov:2004tf}),
\begin{equation}
\label{ab}
a'(\xi) = \frac{1+\mathrm{e}^{-\theta}}{1-\mathrm{e}^{-\theta}}\cdot 
\frac{1}{\pi \xi} \mathrm{V.P.}\int d\xi_1 
\frac{\xi_1 b'(\xi_1)}{\xi_1-\xi}\;,
\end{equation}
where the integral is understood in the sense of principal value.
We use Eq.~\eqref{ab} as an alternative formulation of the 
condition~\eqref{BCI} in our model. 

One also requires that the functions $\varphi_{\mathrm{in}}$, 
$\varphi_{\mathrm{out}}$ have appropriate 
asymptotics. To ensure the finite number of particles in the initial state
we require the incoming wave to be well 
localized, $\varphi_{\mathrm{in}}(\xi)\to0$ as $\xi\to\pm\infty$. 
Besides, as the initial state is an excitation above the vacuum
$\phi=0$, we write $\varphi_{\mathrm{out}}(\xi)\to 0$ at $\xi\to-\infty$. 
On the other hand, as we have 
already discussed, the relevant solution contains the sphaleron at 
$t\to +\infty$.
Thus, $\varphi_{\mathrm{out}}(\xi)\to \pi$ as  
$\xi\to +\infty$. 

The problem~\eqref{b},~\eqref{c},~\eqref{ab} can be solved numerically by
the following iterative method. At each cycle of iterations one starts from
the function\footnote{
At the very first cycle one takes, e.g.,
$u^{(0)}(\xi) = 
\pi/2 + \arctan(\mu \xi)$.} $u = u^{(0)}(\xi)$ and solves Eq.~\eqref{b}
explicitly,
\begin{equation}
b(\xi) = \ln \mathrm{th} 
\left(
  -\frac{\mu}{2}\int_0^{\xi} \cos u^{(0)}(\xi_1) d\xi_1 + \varkappa
\right)\;,
\end{equation}
where $\varkappa$ is an integration constant. Then, integrating numerically 
Eqs.~\eqref{ab},~\eqref{c}, one finds 
$a(\xi)$ and $u(\xi)$ for given value of $\varkappa$. 
Finally, one picks up the
 value of $\varkappa$ such that the function $u(\xi)$ has correct asymptotics
as $\xi\to\pm\infty$.
In this way the improved approximation for $u(\xi)$ is obtained, and a
new cycle 
of iterations begins. After 30 cycles one obtains the solution with precision 
of order $10^{-6}$.

\begin{figure}
\centerline{
\includegraphics[width=0.5\columnwidth]{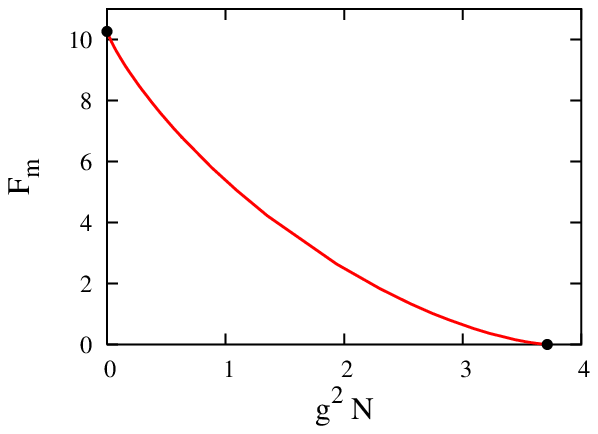}
\includegraphics[width=0.5\columnwidth]{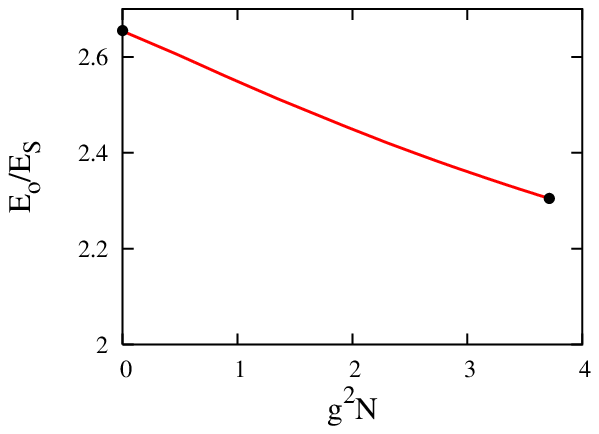}
}
\caption{FIG. 1. 
Suppression exponent (left) and energy of real--time instantons 
(right) versus the number of incoming particles in the model (\ref{L}).} 
\label{fig:1}
\end{figure}

Given the family of real--time instantons, one calculates numerically the 
suppression exponent $F_{\rm m}(\theta)$ via Eq.~\eqref{Fmresult}, and 
also
the energy $E_{\rm o}(\theta)$. To calculate the number of particles it is
convenient to recast Eq.~\eqref{N} in the  
form
\begin{equation}
\tilde{N}\equiv g^2 N = 
-\frac{2}{\mathrm{sh} \theta}\int d\xi~ a'(\xi) b(\xi)\;.
\end{equation}
Functions $F_{\rm m}(\theta)$, $E_{\rm o}(\theta)$ 
and $\tilde{N}(\theta)$ determine the dependence of the suppression 
exponent and the energy on $\tilde{N}$. These are 
plotted in Fig.~\ref{fig:1}.
Taking the limit $\tilde{N}\to 0$  one obtains 
$F_{\rm m} =10.27$, $E_{\rm o}=2.65E_S$. Thus, the semiclassical
suppression factor is $exp(-10.27/g^2)$ at all energies
exceeding $E_{\rm o}=2.65E_S$, and the suppression is even
stronger at lower energies.

In conclusion we summarize our method. To calculate the minimum
suppression exponent $F_{\rm m}$ and the optimal energy $E_{\rm o}$ of
a collision--induced tunneling process, one finds the family of
complex classical solutions, real--time instantons, satisfying  the
boundary conditions~\eqref{BC}. Given the real--time instantons, one
calculates the number of particles $\tilde{N}$, the suppression
exponent $F_{\rm m} (\tilde{N})$, Eqs.~\eqref{N},~\eqref{Fmresult},
and the energy $E_{\rm o}(\tilde{N})$. By itself, the 
quantity $F_\mathrm{m}(\tilde{N})$ provides the lower bound on
$F_\mathrm{m}$, Eq.~\eqref{FleqF}; the limit $\tilde{N}\to 0$
yields $F_{\rm m}$ and $E_{\rm o}$ according to 
Eqs.~\eqref{lim},~\eqref{EleqElN}.

We are indebted to S.~Dubovsky, V.~Rubakov, P.~Tinyakov and F.~Bezrukov
for helpful suggestions. This work has been supported in part by 
RFBR grant 02-02-17398, grant NS-2184.2003.2 and fellowships 
of the ``Dynasty'' foundation (awarded by the Scientific 
board of ICFPM). 
Work of D.~L. has been supported by CRDF award RP1-2364-MO-02 and 
INTAS grant YS03-55-2362. D.L. is grateful to the Universite Libre de
Bruxelles for hospitality.


\begin{thebibliography}{99}
\bibitem{Coleman:1977py}
  S.~R.~Coleman,
  \newblock Phys.\ Rev.\ D {\bf 15}, 2929 (1977) 
  [Erratum-ibid.\ D {\bf 16}, 1248 (1977)];
  \newblock HUTP-78/A004
  {\it Lecture delivered at 1977 Int. School of Subnuclear Physics, 
  Erice, Italy, Jul 23-Aug 10, 1977}
\bibitem{Belavin:1975fg}
  A.~A.~Belavin, A.~M.~Polyakov, A.~S.~Schwartz and Y.~S.~Tyupkin,
  \newblock Phys.\ Lett.\ B {\bf 59}, 85 (1975).
\bibitem{Klinkhamer:1984di}
  N.~S.~Manton, \newblock Phys.\ Rev.\ D {\bf 28}, 2019 (1983).\\
  F.~R. Klinkhamer and N.~S. Manton,
  \newblock Phys. Rev. {\bf D30}, 2212 (1984).
\bibitem{Ringwald:1990ee}
  A.~Ringwald,
  \newblock Nucl. Phys. {\bf B330}, 1 (1990).\\
  O.~Espinosa,
  \newblock Nucl. Phys. {\bf B343}, 310 (1990).
\bibitem{Kuznetsov:1997az}
  A.~N. Kuznetsov and P.~G. Tinyakov,
  \newblock Phys. Rev. {\bf D56}, 1156 (1997).
\bibitem{Bezrukov:2003er}
  F.~Bezrukov, D.~Levkov, C.~Rebbi, V.~Rubakov and P.~Tinyakov,
  \newblock Phys.\ Rev.\ D {\bf 68}, 036005 (2003); 
  \newblock Phys.\ Lett.\ B {\bf 574}, 75 (2003).

\bibitem{Voloshin:1993dk}
  M.~B.~Voloshin,
  \newblock Phys.\ Rev.\ D {\bf 49}, 2014 (1994).
\bibitem{Rubakov:1994hz}
  V.~A.~Rubakov and D.~T.~Son,
  \newblock Nucl.\ Phys.\ B {\bf 424}, 55 (1994).
\bibitem{Zakharov:1990xt}
  V.~I.~Zakharov,
  \newblock Nucl.\ Phys.\ B {\bf 353}, 683 (1991).\\
  G.~Veneziano,
  \newblock Mod.\ Phys.\ Lett.\ A {\bf 7}, 1661 (1992).\\
  M.~Maggiore and M.~A.~Shifman,
  \newblock Nucl.\ Phys.\ B {\bf 371}, 177 (1992).
\bibitem{Levkov:2004tf}
  D.~G.~Levkov and S.~M.~Sibiryakov,
  arXiv:hep-th/0410198.
\bibitem{Rubakov:1992ec}
  V.~A.~Rubakov, D.~T.~Son and P.~G.~Tinyakov,
  \newblock Phys.\ Lett.\ B {\bf 287}, 342 (1992).
\bibitem{Tinyakov:1991fn}
  P.~G.~Tinyakov,
  \newblock Phys.\ Lett.\ B {\bf 284}, 410 (1992).\\
  A.~H.~Mueller,
  \newblock Nucl.\ Phys.\ B {\bf 401}, 93 (1993).
\bibitem{Bonini:1999kj}
  G.~F.~Bonini, A.~G.~Cohen, C.~Rebbi and V.~A.~Rubakov,
  \newblock Phys.\ Rev.\ D {\bf 60}, 076004 (1999).
\bibitem{Bezrukov:2003tg}
  F.~Bezrukov and D.~Levkov,
  \newblock J.\ Exp.\ Theor.\ Phys.\  {\bf 98}, 820 (2004)
  [Zh.\ Eksp.\ Teor.\ Fiz.\  {\bf 125}, 938 (2004)].
\bibitem{Tinyakov:1992dr}
  P.~G.~Tinyakov,
  Int.\ J.\ Mod.\ Phys.\ A {\bf 8} (1993) 1823.
\bibitem{Rebbi:1996qt}
  C.~Rebbi and R.~J.~Singleton, 
  \newblock hep-ph/9706424.
\bibitem{Kane}
  C.~L.~Kane and  M.~P.~A.~Fisher,
  \newblock Phys.\ Rev.\ Lett.\  {\bf 68}, 1220 (1992).
\bibitem{Fazio}
  R.~Fazio, K.~H.~Wagenblasta, C.~Winkelholza and G.~Sch\"{o}n,
  \newblock Physica {\bf B222}, 364 (1996).
\end{thebibliography}
\end{document}